\documentclass[fleqn,usenatbib]{mnras}

\usepackage{newtxtext,newtxmath}
\usepackage[T1]{fontenc}
\usepackage{ae,aecompl}
\usepackage{graphicx}	
\usepackage{amsmath}	
\usepackage{amssymb}	
\usepackage{xspace}
\usepackage{xcolor}

\newcommand{\Nbody}{$N$-body\xspace}
\newcommand{\nbodysix}{{\sc{nbody6}}\xspace}
\newcommand{\myosotis}{{\sc{MYOSOTIS}}\xspace}
\newcommand{\gandalf}{{\sc{gandalf}}\xspace}
\newcommand{\starfinder}{{\sc{starfinder}}\xspace}
\newcommand{\percent}{\,{\rm per\,cent}\xspace}
\newcommand{\Msun}{\,\mathrm{M}_{\odot}}
\newcommand{\magnitude}{\,\mathrm{mag}}
\newcommand{\logten}{\,\mathrm{log_{10}}}
\newcommand{\Ks}{K_\mathrm{s}}
\newcommand{\pc}{\,\mathrm{pc}}

\newcommand{\secref}[1]{Section~\ref{#1}}
\newcommand{\figref}[1]{Fig.~\ref{#1}}
\newcommand{\tabref}[1]{Table~\ref{#1}}
\newcommand{\equref}[1]{equation~\eqref{#1}}

\title[Make Your Own Synthetic ObservaTIonS]{A Code to Make Your Own Synthetic ObservaTIonS (MYOSOTIS)}

\author[Z. Khorrami et al.]{
	Zeinab Khorrami$^{1}$\thanks{E-mail: KhorramiZ@cardiff.ac.uk},
	Pouria Khalaj$^{2}$,
	Anne S. M. Buckner$^{3}$,
	Paul C. Clark$^{1}$,
	\newauthor
	Estelle Moraux$^{2}$,
	Stuart Lumsden$^{3}$,
	Isabelle Joncour$^{2,4}$,
	Ren\'e D. Oudmaijer$^{3}$,
	\newauthor
	Ignacio de la Calle$^{5}$,
	Jos\'e M. Herrera-Fernandez$^{5}$,
	Fr\'ed\'erique Motte$^{2}$,
	\newauthor
	Jos\'e Manuel Blanco$^{5}$
	and
	Luis Valero-Martin$^{5}$
	\\
	$^{1}$School of Physics and Astronomy, Cardiff University, The Parade, CF24 3AA, UK\\
	$^{2}$Universit\'e Grenoble Alpes, CNRS, IPAG, 38000 Grenoble, France\\ 
	$^{3}$School of Physics and Astronomy, University of Leeds, Leeds LS2 9JT, U.K.\\
	$^{4}$Department of Astronomy, University of Maryland, College Park, MD 20742, USA\\
	$^{5}$Quasar Science Resources, S.L., Edificio Ceudas, Ctra. de La Coru\~na, Km 22.300, 28232, Las Rozas de Madrid, Madrid, Spain
}

\date{Accepted XXX. Received YYY; in original form ZZZ}
\pubyear{2018}
\begin{document}
	\label{firstpage}
	\pagerange{\pageref{firstpage}--\pageref{lastpage}}
	\maketitle
	
	\begin{abstract}
    We introduce our new code \myosotis (Make Your Own Synthetic ObservaTIonS) which is designed to produce synthetic observations from simulated clusters. The code can synthesise observations from both ground- and spaced-based observatories, for a range of different filters, observational conditions and angular/spectral resolution. 
    In this paper, we highlight some of the features of \myosotis, creating synthetic observations from young massive star clusters.         
	Our model clusters are simulated using \nbodysix code and have different total masses, half-mass radii, and binary fractions. The synthetic observations are made at the age of 2\,Myr with Solar metallicity and under different extinction conditions. For each cluster, we create synthetic images of the Hubble Space Telescope (HST) in the visible (WFPC2/F555W) as well as Very Large Telescopes (VLT) in the nearIR (SPHERE/IRDIS/$\Ks$). 
    We show how \myosotis can be used to look at mass function (MF) determinations. For this aim we re-estimate stellar masses using a photometric analysis on the synthetic images. The synthetic MF slopes are compared to their actual values. Our photometric analysis demonstrate that depending on the adopted filter, extinction, angular resolution and pixel sampling of the instruments, the power-law index of the underlying MFs can be shallower than the observed ones by at least $\pm0.25$\,dex which is in agreement with the observed discrepancies reported in the literature, specially for young star clusters.
	\end{abstract}
	
	
	\section{Introduction}

We have been using N-body models to study the physics of star clusters since Van Albada \citep{vanalbada1968}. Such simulations, have been used to look at stellar cluster core oscillations \citep{heggie2009, giersz2009, hurley2012}, stellar collisions \citep{chatterjee2009}, merging of star clusters \citep{priya2016}, the evolution of multiple systems \citep{hurley2002} and substructures \citep{allison2009}, and the phenomenon of mass segregation and its role in cluster evolution (e.g. \citealt{simon2010}).
Through this work, the community has built up a picture of how clusters evolve (e.g. \citealt{kalirai2010}), and how this may effect the initial mass function (IMF; e.g. \citealt{kroupa2001}). They have also been used to place constraints on the star formation process (e.g. \citealt{parker2013}) and how cluster dynamics can affect the stability of planetary systems (e.g. \citealt{cai2017}).  
  
The results from N-body modeling have been used to help interpret the results from observational studies. For example the evolution of mass function (MF) slope in globular clusters \citep{baumgardt2008}, relation between the MF slope and the core radius of the star cluster \citep{demarchi2007} and the debate on the observed mass-segregation (primordial, dynamical or observational bias) and its origin in star clusters \citep{dominguez2017, parker2016, bastian2010,espinoza2009}. 

Although comparisons between observations and simulations are often made directly to the N-body results, such a comparison is dangerous since:
	\begin{enumerate}
		\item Most of the young star clusters (YSCs) contain hot and massive stars, which can mask the faint low-mass stars. This leads to an underestimation of the number of low-mass stars which makes the observed MF steeper than the underlying MF at the low-mass end.
		
		\item YSCs are immersed in their natal cloud \citep{lada2003,simon2010} meaning that stellar members suffer from extinction which varies from point to point. This means, that applying a constant value of extinction to the entire stellar population inside the cluster, leads to an incorrect estimation of stellar masses, and consequently a deformed MF. 
		
		\item Individual members are not fully resolved for most known YSCs due to their typically large distances. This is especially important for unresolved multiple stars (e.g. binaries) which can affect the measured low- and high-mass slopes of the MF (e.g. \citealt{malkov2001, khalaj2013})  
	\end{enumerate}

	Considering all the aforementioned observational difficulties, we need to observe YSCs with better angular resolutions and high contrast imaging and preferentially at longer wavelengths. Furthermore, numerical simulations and models of YSCs are dictated and evolve according to observations, but the comparison of the two is far from straightforward. In particular, we always need to take an intermediate step to create synthetic observations from the simulations first, and only then the comparison with the observations is sensible. 

In this paper we introduce our code \myosotis (Make Your Own Synthetic ObservaTIonS), a tool for creating synthetic observational data which produces the imaging and spectroscopic data for space- and ground-based telescopes. Using \myosotis one can change the angular resolution of the observing instrument, pixel sampling of the detector, extinction and the atmospheric conditions. These factors significantly affect the photometric analysis of the individual stars detected in the field of view (FOV), especially in crowded field images like star clusters. 
 
	\myosotis enables us to create synthetic images/spectra of telescopes such as the Hubble Space Telescope (HST), Very Large Telescopes (VLT) and Gaia, from the \Nbody simulations, to be compared with real data of the aforementioned telescopes. Moreover, Our tool can be used with custom configurations, meaning, that it can replicate the observations of a wide variety of instruments.   
		
	\myosotis has been developed as part of the StarFormMapper\footnote{\url{http://sfm.leeds.ac.uk}} (SFM) project, which aims to study massive stars and star cluster formation using Gaia and Herschel data. 
		To this end, we aim to examine how synthetic observations from different telescopes produce different results on the MF of YSCs, and whether it is possible to attribute the observed discrepancy in the MFs of YSCs to different observational conditions. 
		
		This paper has two main parts: 1) detailed description of how \myosotis works (\secref{sec:myosotis}) and 2) examples of the application of the code (\secref{sec:applications}). 
		In the latter part we use \myosotis to create synthetic HST/WFPC2 and VLT/SPHERE images of YSCs in the visible ($V$-band) and nearIR ($\Ks$-band) from \Nbody simulations. 
		The details of the \Nbody simulations are given in \secref{sec:nbody}. The synthetic observational data which are created by \myosotis are explained in \secref{sec:scenes}. 
		In \secref{sec:analysis} we explain the photometric method that we used to analyze the synthetic images, extract stellar sources and finally estimate their masses. One example of synthetic spectroscopic data is given in \secref{sec:spectroscopy}.
		A discussion and summary of the results is presented in \secref{sec:summary}.

		\section{MYOSOTIS: Make Your Own Synthetic ObservaTIonS}\label{sec:myosotis}
		
		This code creates synthetic imaging and spectroscopic data of space and ground-based telescopes as well as custom (user-defined) instruments within any FOV.
		The stellar and interstellar medium information (position, mass, velocity, metallicity, age) should be provided by the user.
		The user can choose different filters from a list (see {\url{http://svo2.cab.inta-csic.es/theory/fps/}}) or define a new filter, to suit the observational instrument that they want to mimic.
		The observing conditions, i.e. seeing, Strehl-Ratio (SR), detector's pixel scale of a given instrument, FOV, observer's line-of-sight and finally the angular resolution of the telescope can be defined in \myosotis. Since most of the instruments can not achieve their theoretical optimum resolution ($\sim \lambda$/Diameter), the user can also define their own resolution.
		The estimated flux of stellar sources spreads on the detector using a 2D point spread function (PSF) whose full width at half maximum (FWHM) is equal to the resolution. The user can choose a Gaussian distribution or an Airy pattern for the PSF of stellar sources. 
        The extinction can be applied on the output data, knowing the column density of the gas in front of each source. 
		This extinction could be uniform, patchy, or taken from a full 3D smoothed particle hydrodynamics (SPH) simulation data.
        
		\subsection{Stellar evolutionary and atmosphere models} \label{sec:EAmodels}
		One of the input files for \myosotis is the information on stellar positions, velocities, masses, ages and metallicities.
		For each star, according to its age, metallicity and mass, \myosotis finds the closest stellar parameters, i.e. effective temperature ($T_{\rm eff}$), surface gravity (log$\,$g) and luminosity (log$\,L$), using the grids of {\sc{parsec}} \footnote{We have used {\sc{parsec}} isochrones (vesrion 1.2S) and CMD 3.0 web interface avialable at \url{http://stev.oapd.inaf.it/cgi-bin/cmd}} evolutionary models \citep{parsec2012,chen2014,chen2015,tang2014}. {\sc{parsec}} has a complete theoretical library that includes the latest set of stellar phases from pre-main sequence to main sequence, covering stellar masses from $0.09$ to $350\Msun$ and ages between 0.1\,Myr up to 10.1\,Gyr.
		
		After finding $T_{\rm eff}$ and log$\,$g for each star, \myosotis finds the closest stellar atmosphere model that is, the full spectral energy distribution (SED), that fits the given metallicity, $T_{\rm eff}$ and log$\,$g of that star. \myosotis uses the grids of 
		{\sc{NextGen2}} atmosphere models for very low-mass stars and brown dwarfs with $900\,K < T_{\rm eff} < 3400\,K$ covering log$\,$g from 3.5 to 6.0 \citep{nextgen97,nextgen99}, and {\sc{atlas9}} {\sc{kurucz}} {\sc{odfnew/nover}} atmosphere models \citep{kurucz97} for $3500\,K < T_{\rm eff} < 50000\,K$ covering log$\,$g from 0.0 to 5.0 for solar metallicity.
		
\begin{figure}
			\includegraphics[width=\columnwidth]{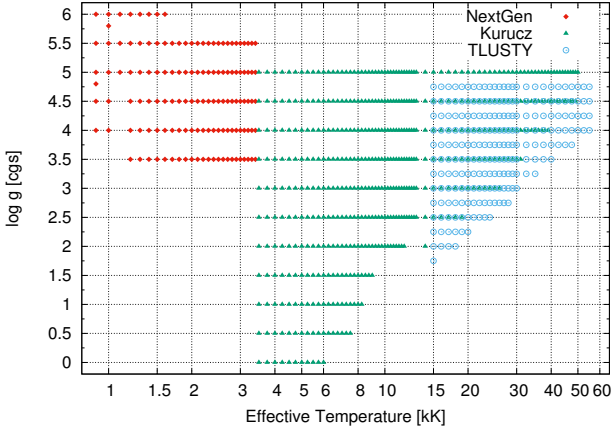}    
			\caption{The stellar parameters of the atmosphere models used in \myosotis.}
			\label{fig:seds}
		\end{figure}

		Users can choose a specific atmosphere model for hot and massive O- and B-type stars by setting the OB treatment parameter to `yes' (i.e. {\bf{OBtreatment=`yes'}}). In this case, for stars with $T_{\rm eff} > 15000\,K$, \myosotis uses the grids of {\sc{tlusty}} \footnote{\url{http://nova.astro.umd.edu/Tlusty2002/tlusty-frames-cloudy.html}} atmosphere models \citep{tlusty} for B-type \citep{tlustyB} and O-type \citep{tlustyO} stars. {\sc{Tlusty}} grids, cover $T_{\rm eff}$ from 15000\,K up to 55000K and log$\,$g from 1.75 up to 4.75. \figref{fig:seds} shows the stellar parameters covered by these atmosphere models for solar metallicity. After selecting the appropriate SED for each star, \myosotis will estimate the stellar flux and extinction in a given filter, according to its distance from the observer. 
		In addition to the aforementioned models for stellar evolution and atmospheres, users can define their own customized models and any attenuating dust that has been prescribed by the user. 
		
		\subsection{Bolometric Corrections and Extinction}
		To estimate the bolometric correction (BC) of stars in different filters, we used the method explained in \cite{girardi2002}, i.e. for a given filter BC is given by 
		
		\begin{equation}\label{eq:bc}
		\begin{aligned}
		BC_{S_\lambda} =& M_{\rm bol,\odot} - 2.5 log[4 \pi (10 pc)^2 \sigma T_{eff}^4/L_\odot] \\
		& + 2.5 log[\frac{\int_{\lambda_1}^{\lambda_2} \lambda F_\lambda 10^{-0.4 A_\lambda} S_\lambda d\lambda}{\int_{\lambda_1}^{\lambda_2} \lambda f^0_\lambda S_\lambda d\lambda}] - m^0_{S_\lambda}
		\end{aligned}
		\end{equation}
		
		In this equation, $M_{\rm bol,\odot}=4.83$ and $L_\odot=3.828\times 10^{33} \rm erg/\rm s$ {\footnote{"Sun Fact Sheet" \url{https://nssdc.gsfc.nasa.gov/planetary/factsheet/sunfact.html}}}. $A_\lambda$ is the extinction at wavelength $\lambda$ and $S_\lambda$ is the filter transmission curve corresponding to the interval $[{\lambda_1},{\lambda_2}]$.
		$F_\lambda$ is the stellar intrinsic spectra at wavelength $\lambda$ which is provided by the atmosphere model for a given $T_{eff}$, log$\,$g and metallicity.
		$f^0_\lambda$ is the reference spectra of Vega at the Earth surface \footnote{\url{http://basti.oa-teramo.inaf.it/BASTI/MAG_ML/Vega.sed}} that produces a known apparent magnitude $m^0_{S_\lambda}$ in different wavelengths. Vega has $V = 0.034$ mag (3670 Jy for $V=0$), and all colours are equal to 0. 
		For the Vega spectrum we used synthetic {\sc{atlas9}} model, with $T_{eff} = 9550$\,K, log$\,g = 3.95$ and $[\rm{M}/\rm{H}] = −0.5$ provided by \citet{kurucz97}.

		\subsection{Cloud column density}\label{app:column}
       The column density of the cloud in front of each stellar source (uniform or patchy) can be given directly by the user or \myosotis can calculate it using the SPH data if it is available. In the case of SPH data, the gas cloud can be located anywhere around the stellar sources as well as anywhere within the line-of-sight of the observer. 
        \myosotis then calculates the cloud column density in front of each star (in the line-of-sight of the observer). The user should provide the cloud information (cloud particles positions, mass and smoothing lengths). This information is the standard output of the SPH simulations. In SPH, particle properties are smoothed over a length scale, $h$, called the smoothing length, using a weighting function, $W(r, h)$, called the kernel function. \myosotis uses M4 cubic spline kernel function \citep{ML85}, shown in \equref{eq:kernel}.
		
		\begin{equation}\label{eq:kernel}
		W(r/h) = \frac{1}{\pi h^3}
		\begin{cases}
		1- \frac{3}{2} (r/h)^2 + \frac{3}{4} (r/h)^3 & {0 \leq r/h \leq 1} \\
		\frac{1}{4} (2 - r/h)^3 & {1 \leq r/h \leq 2} \\
		0 & {2 \leq r/h}
		\end{cases}
		\end{equation}
		
\begin{figure}
			\includegraphics[width=\columnwidth]{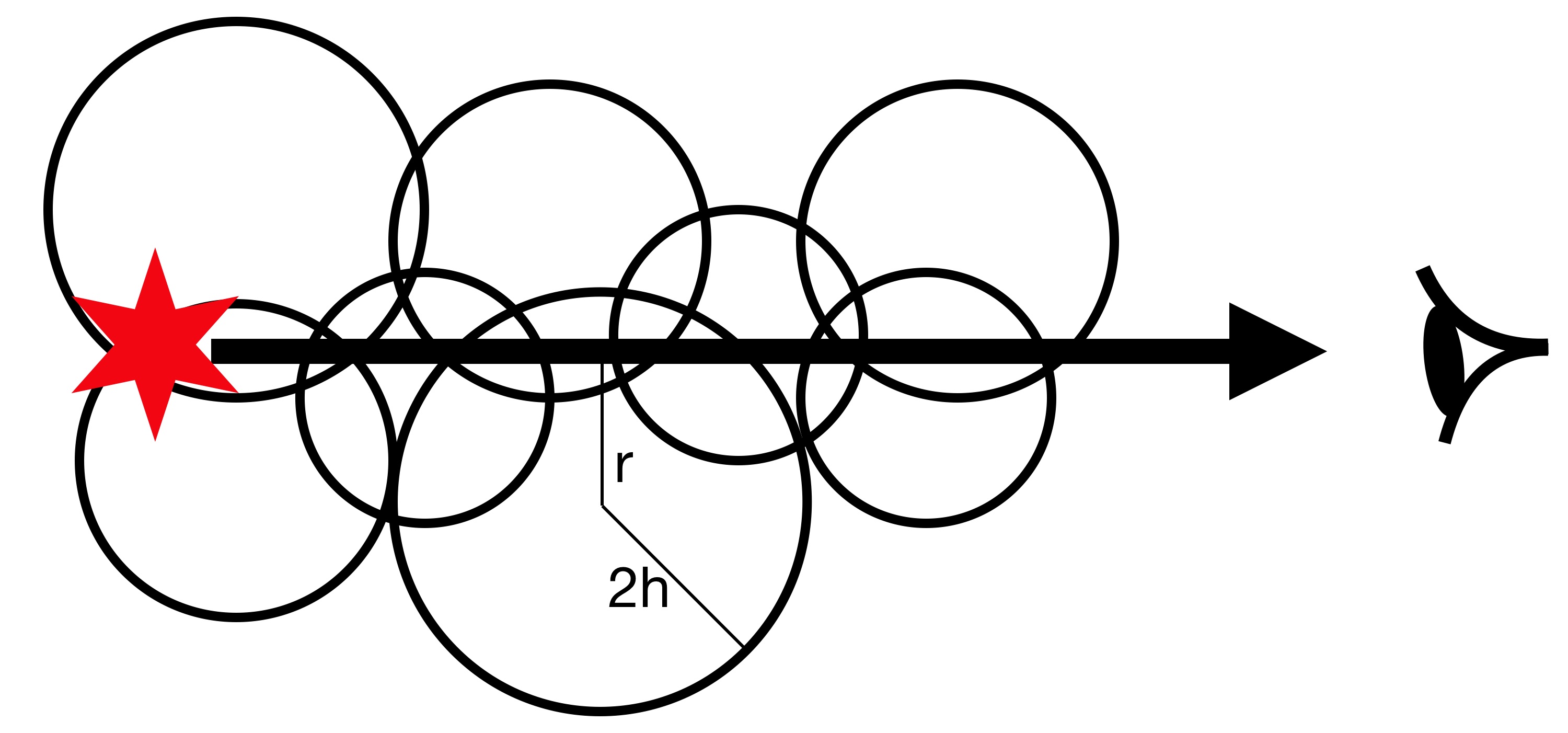} 
			\caption{The cloud particles distributed in front of an stellar source. Each cloud particle has a smoothing length ($h$) and the column density is calculated as a function of $r$ for the particles lying in the line-of-sight using the kernel function given in \equref{eq:kernel}.}
			\label{fig:cloud}
		\end{figure}

		\figref{fig:cloud} is a schematic representation of the cloud particles distributed in front of a stellar source. \myosotis detects the cloud particles which are located in the line-of-sight of the observer and the stellar source. Each cloud particle has a smoothing length and a mass. The column density of the cloud can be calculated using the kernel function (\equref{eq:kernel}), as a function of distance from the centre of each cloud particle and its mass. After estimating the column density in front of each star, we use the relation between optical extinction (A$_V$) and Hydrogen column density $N_{\rm H} [{\rm cm}^{-2}]$ from \cite{GO09}, i.e. 
		\begin{equation}
		N_{\rm H} [{\rm cm}^{-2}] = (2.21 \pm 0.09) \times 10^{21} A_V [\magnitude].
		\end{equation}
		
		\subsection{Extinction}\label{app:ext}
		The extinction for each stellar source, can be calculated using two different methods:
		
		\begin{itemize}
			\item {\bf{Fmodel}}: The code uses a function to calculate extinction in a given wavelength knowing optical extinction values (A$_V$ and R$_V$).
			This function uses the average extinction curve in the optical-through-IR range (0.125 - 3.333 $\mu$m) which is reproduced with a cubic spline and a set of anchor points from \cite{fitzpatrick1999}.
			
			\item {\bf{Dmodel}}: The code uses synthetic extinction curves \footnote{\url{www.astro.princeton.edu/~draine/dust/dustmix.html}} from \cite{drainea, draineb, drainec, LD01, WD01}. 
			Extinction, absorption, albedo, $<{\rm cos}(\theta)>$, and $<{\rm cos}^2(\theta)>$ have been calculated for wavelengths from 1 cm (30 GHz) to $1$ Angstrom (12.4 keV), for selected mixtures of carbonaceous grains and amorphous silicate grains.
			These models cover three values of R$_V$, 3.1, 4.0 and 5.5.
		\end{itemize}
		
		The required input for both extinction models is A$_V$. In \secref{app:column} we explain how the code estimates A$_V$ in the line-of-sight of each star.

		\subsection{Observing condition}
		As explained earlier, \myosotis provides a customized list of telescope filters. In addition to the filter transparency, \myosotis needs to know the pixel-scale and the angular resolution of the observing instrument. If the telescope is ground-based ({\bf{Adaptiveoptics=`yes'}}), then the user should also provide the atmospheric conditions e.g. SR and the seeing values. If the ground-based telescope does not have any adaptive optics and its optimum resolution is poorer than seeing, user can simply choose an angular resolution equal to the seeing. 
        The distance to the center of mass of the object (e.g. star cluster) and the FOV should be also provided by the user. Note that all stars will not have the same distance from the observer as \myosotis calculates the exact distance of each star, according to its 3D position. It is also possible to change the orientation of the object according to the observer's line-of-sight. Our tool can apply the Doppler shift on the spectra of each star, according to its 3D velocity. Moreover, one can chose different values of signal-to-noise (SNR) for the faintest star in the FOV. In this case \myosotis will provide an extra FITS image with noise.

		\section{Applications}\label{sec:applications}
		
		To show some of the basic applications of \myosotis, we simulated four star clusters using  the publicly available code \nbodysix \citep{nbody6}. The initial conditions for these simulations is given in the following section.
		We generated synthetic observational data from these star clusters, using \myosotis. Then we analyze these data step by step using standard photometric methods to:
		
		1) extract stellar sources in each image 
		
		2) find common stars between the data sets in different wavelengths 
		
		3) fit isochrones to the CMD in order to estimate the age of the star cluster
		
		4) estimate stellar masses in different filters in a given age
		
		5) apply artificial stellar source recovery tests, to estimate the completeness as a function of stellar mass
		
		6) plot mass functions and find the derived slope of the IMF in our synthetic images 
		
		For one of our simulations (Sim5), we embedded the stars within a cloud of SPH particles, to demonstrate the ability of \myosotis to treat patchy extinction. Details of the cloud are given below in \secref{sec:scenes}.
		
		For one of our simulations (Sim1) we also show a small region from the centre of cluster, and compare the true positions of the stars with those derived from the photometry. In addition, we use this region to show how \myosotis can be used to investigate blending in stellar spectra.
		
		\subsection{{\it{N}}-body simulations of star clusters}\label{sec:nbody}
		We use \nbodysix to simulate the dynamical and stellar evolution of four different clusters. The clusters are set up using {\sc{mcluster}} \citep{kupper11}. \tabref{tab:simsIC} summarizes the initial conditions of the simulated clusters. As shown in the table, the simulated clusters differ in total initial mass ($10^4$ and $10^5$ M$_\odot$), half-mass radius (0.5 and 0.8 pc) and binary fraction (0, 30 and 50 \percent). 
		We have used a Plummer model \citep{plummer1911} for the initial mass density profile of these clusters. The initial mass function (IMF) of stellar populations are taken from the distribution given by \citet{kroupa2001}, with a mass range of $0.1-150\Msun$.

		The clusters are initially in virial equilibrium and there is no initial mass segregation. 
		
		For simulations with an initial binary population, the adopted algorithm (in {\sc{mcluster}}) for the pairing of primary and secondary components is as follows. The stars, with a \citet{kroupa2001} IMF, are split into two mass ranges by introducing a mass threshold of $5.0\Msun$. Stars whose mass is below or above this threshold are only paired randomly with stars which belong to the same mass range. This is in rough agreement with the findings of \citet{Kobulnicky07}.
		The period distribution of binaries for the low mass population ($m<5.0\Msun$) was obtained using the period distribution of \citet{kroupa1995a}, and for massive stars ($m>5.0\Msun$) it is based on the distribution reported by \citet{Sana11}. The semi-major axis distribution is obtained from the aforementioned period distributions. The eccentricity ($e$) for low-mass binaries is drawn from a thermal eccentricity distribution, i.e. $f(e)=2e$ (e.g. \citealt{Duquennoy91}; see also \citealt{Kroupa08}), whereas for the high-mass binaries it is from the \citet{Sana11} eccentricity distribution.
		
		\begin{table}
			\centering
			\caption{Initial conditions of the simulated clusters generated using {\sc{mcluster}}. Sim5 is the same as Sim1 except that in the generation and the analysis of the Sim5 synthetic images, the extinction due to a natal cloud is also taken into account.}
			\label{tab:simsIC}
			\begin{tabular}{lccc} 
				\hline
				ID & $M_{\rm tot}$ & $R_{\rm h}$ &Binary Fraction\\
				&[M$_\odot]$& [pc]  & $\%$ \\
				\hline
				Sim1 & $10^4$ & 0.5 & 0 \\
				Sim2 & $10^4$ & 0.5 & 50\\
				Sim3 & $10^5$ & 0.8 & 0  \\
				Sim4 & $10^5$ & 0.8 & 30\\
				Sim5 (Sim1+gas)& $10^4$ & 0.5 & 0  \\
				\hline
			\end{tabular}
		\end{table}
		
		\subsection{Generating Synthetic Images}\label{sec:scenes}
		For each simulated cluster we took one snapshot of the \Nbody simulations at the age of 2\,Myr and created the synthetic observations using \myosotis. The clusters are located at the distance of R136 in the Large Magellanic Cloud (LMC, 50\,kpc from \citealt{lmcdistance}). Although R136 has a metallicity index of $[\rm{M}/\rm{H}] = -0.5$ as appropriate for LMC \citep{dufour1984}, we adopt a Solar metallicity for all simulated clusters, as the current version of \myosotis is limited to only Solar metallicities at present {\footnote{Later versions will include the option to tailor metallicity values.}}. However, this will not affect our results since for the analysis of the photometric data we use the same metallicity that we use for the generation of the synthetic images. Thus, the general trend of our results (flattening of the observed MF; see \secref{sec:analysis}) will not change as a function of metallicity.  
		
\begin{figure}
			\includegraphics[width=\columnwidth]{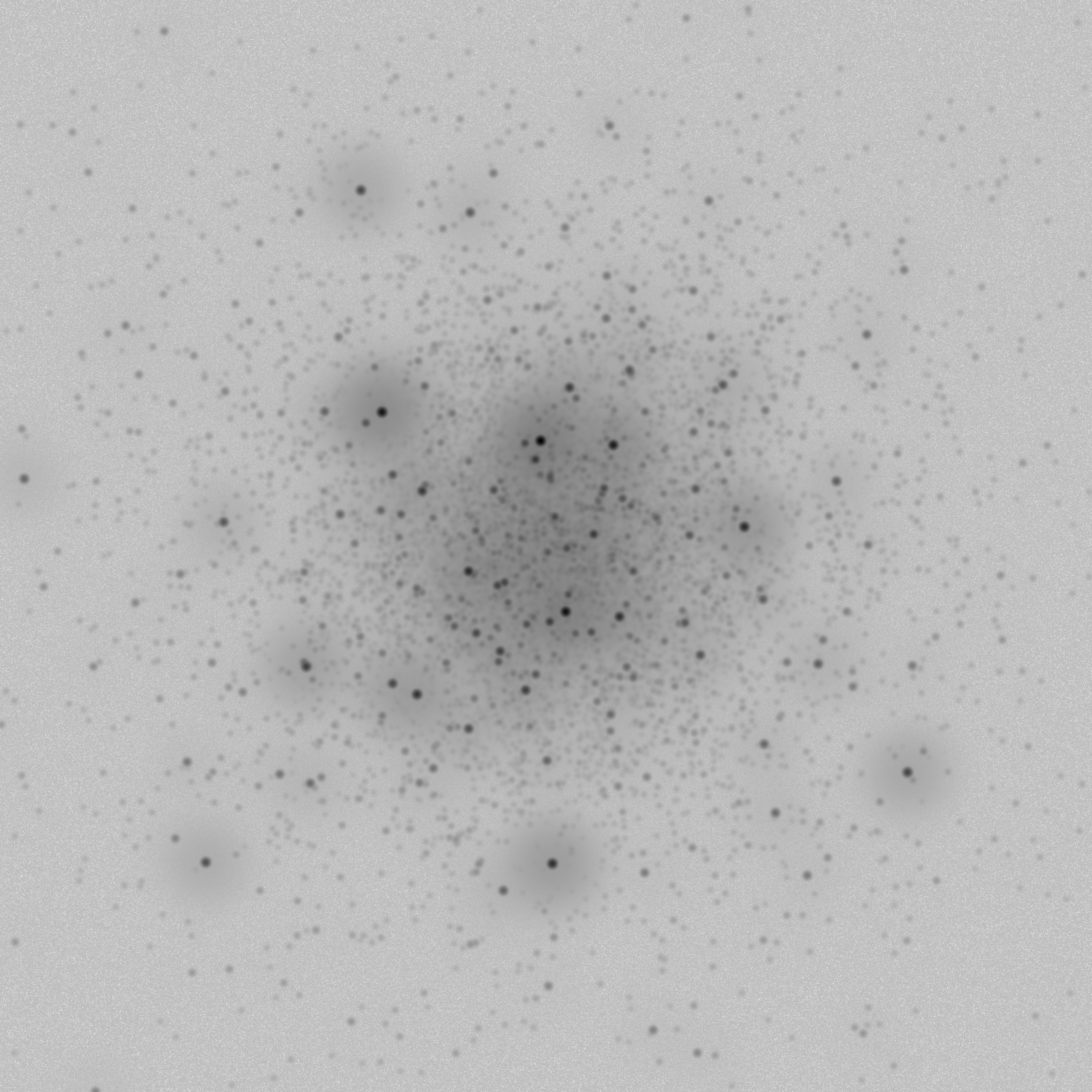}\\
			\includegraphics[width=\columnwidth]{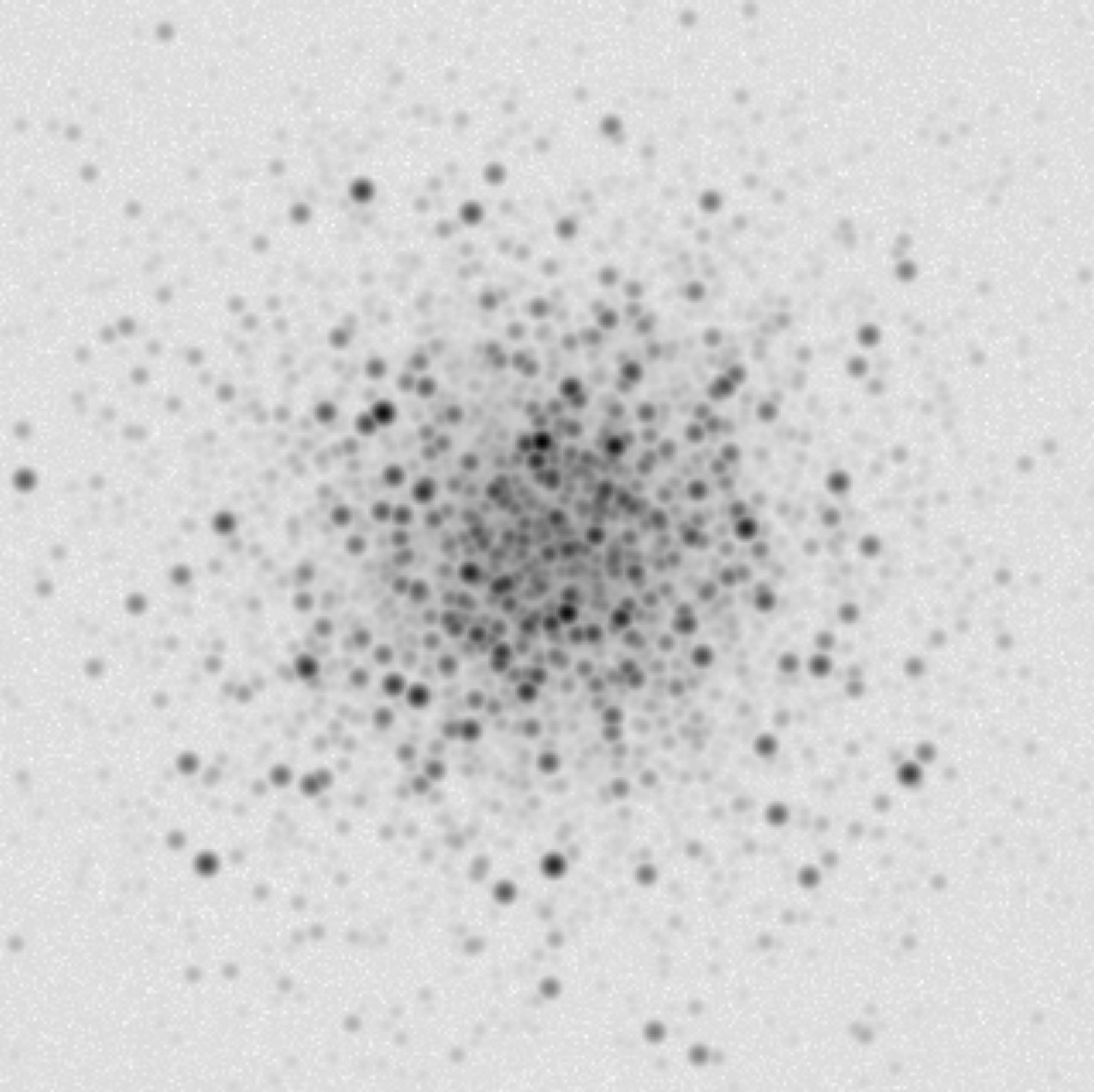}  
			\caption{Synthetic images created from an \Nbody simulation (Sim1 in \tabref{tab:simsIC}) at the age of 2\,Myr. Top: VLT/SPHERE/IRDIS image in nearIR ($\Ks$) with SR=0.75 and seeing=0.8". Bottom: HST/WFPC2 in visible (F555W). The FOV of images is $16" \times 16"$, covering $4\pc \times 4\pc$.}
			\label{fig:im1}
		\end{figure}

		The synthetic images have an FOV of $16" \times 16"$ which corresponds to $4\pc \times 4\pc$. All HST/WFPC2 images have a pixel scale of 50\,mas and an angular resolution of 110\,mas in the visible (F555W filter). For VLT, we simulated SPHERE/IRDIS images in the nearIR ($\Ks$) with a pixel sampling of 12.25\,mas and an angular resolution of 64\,mas. For the atmospheric models of O- and B-type stars in the FOV, we have chosen the OB treatment option (see \secref{sec:myosotis}) in \myosotis. For VLT images, we considered the atmospheric condition, SR to be 0.75 and a seeing halo of 0.8 arcsec. In all the images we applied the shot noise of the sky such that the SNR is 2 for the faintest star in the FOV.
		\figref{fig:im1} shows the simulated images of Sim1 in the nearIR, IRDIS/$\Ks$ (top) and in the visible, WFPC2/F555W (bottom).

\begin{figure}
			\includegraphics[width=\columnwidth]{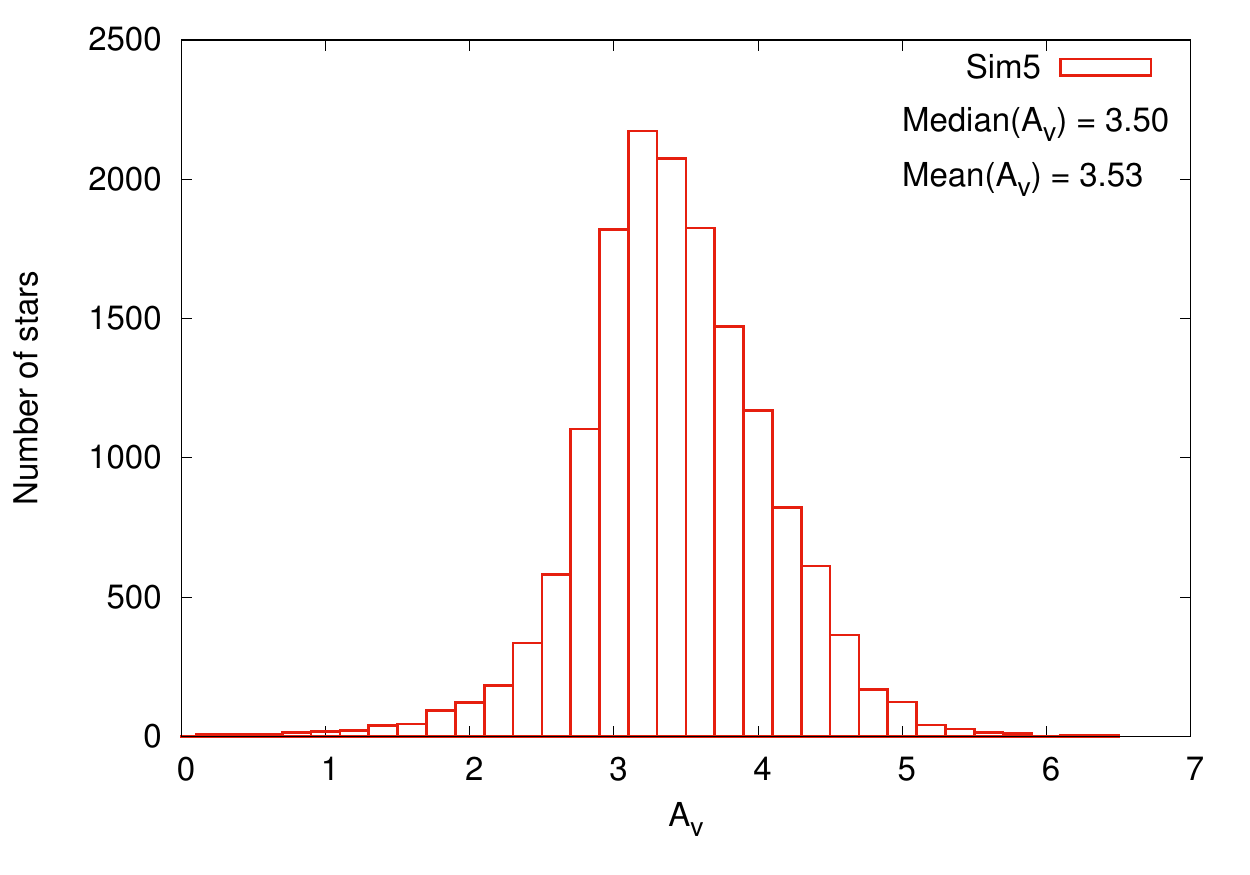}
			\caption{
				The histogram of extinction (A$_v$) in front of each stellar sources in Sim5 simulation. See \tabref{tab:simsIC} for the initial conditions of the simulations.}
			\label{fig:ext}
		\end{figure}

		For simulations Sim1 to Sim4 we did not consider any extinction, meaning, that there is no gas nor dust in the line-of-sight connecting the observer to the stars. This enables us to examine the sole effect of angular resolution on the MF without being concerned about extinction. Sim5 cluster is embedded in its natal cloud which is homogeneous. The center of the cloud is located in the center of the cluster.
        We generated the cloud using the SPH code \gandalf \citep{hubber2016,hubber2018}. The cloud contains $10^5$ SPH particles with smoothing lengths estimated by \gandalf. 
		The total mass of the cloud is $4\times10^3\Msun$ which is $\sim40\percent$ of the total mass of the star cluster. 
	    \figref{fig:ext} shows the histogram of the extinction in front of each stellar source in Sim5. The average value of A$_V$ is 3.5 but depending on the position of the stars in the cloud it varies between $0.2-5.8\magnitude$. This value of $A_V$ is small compare to the $A_V$ of the Galactic young star clusters. As an example, NGC3603 has measured $A_V \sim 4.5$ \citep{khorrami16} and Westerlund 1 has $A_V \sim 11.4$ \citep{damineli2016}.
	    See \secref{app:ext} and \secref{app:column} for more information on how \myosotis estimates column density and extinction in front of each star, using the mass and smoothing length of cloud particles provided by \gandalf.
		
		\subsection{Photometric Analysis And MF Determination}\label{sec:analysis}
\begin{figure}
			\includegraphics[trim=80 0 80 0,clip,width=\columnwidth]{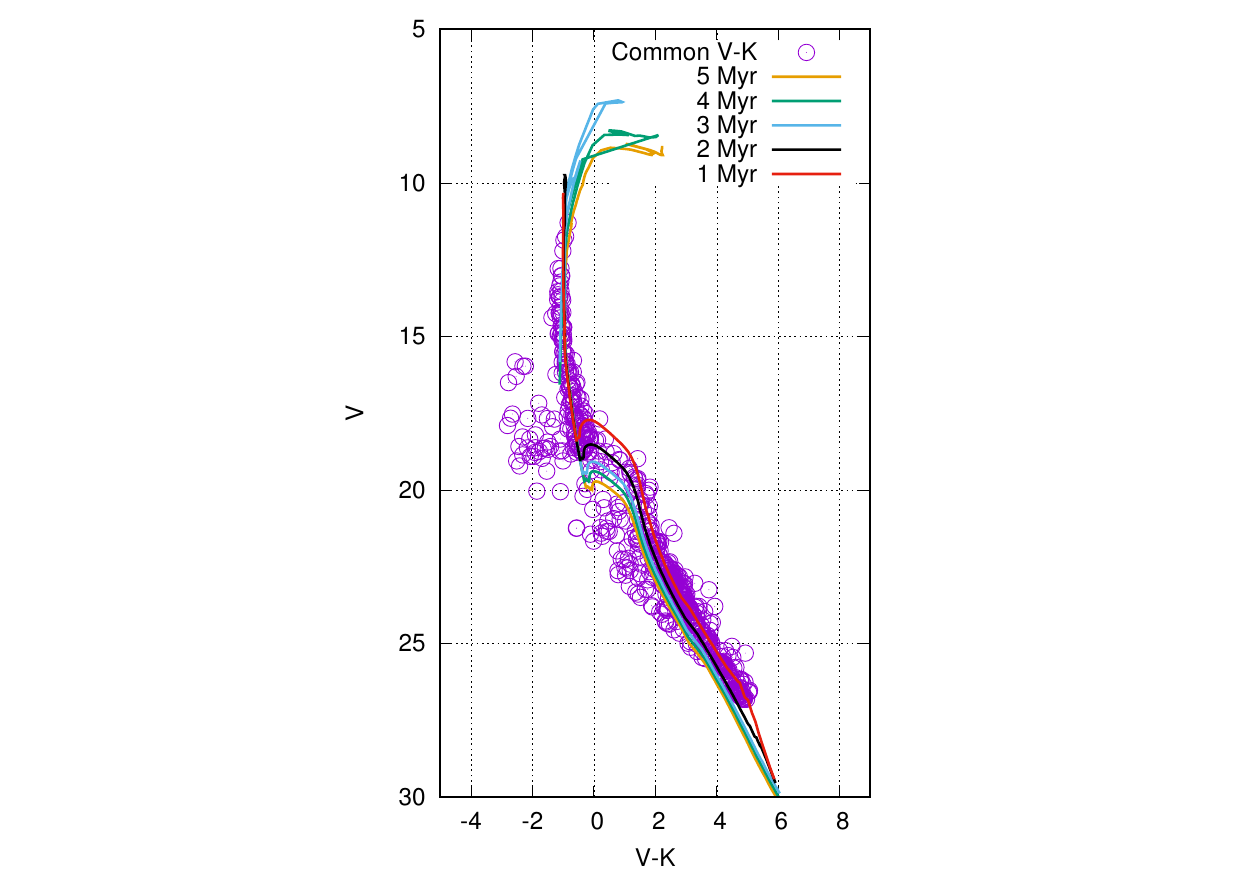}	
			\caption{Color Magnitude Diagram of Sim1 within HST/F555W and IRDIS/$\Ks$ filters.
				Violet circles shows the common sources (747) detected between two images.
				Solid lines are the {\sc{parsec}} isochrones at different ages. }
			\label{fig:cmd}
		\end{figure}

		We used \starfinder \citep{starfinder} to extract the stellar sources from the synthetic images. \starfinder is a suitable code for the deep analysis of stellar fields, designed for AO images with high and low SR.
		The threshold for the photometry in our analysis was chosen to be $4\sigma$ above the sky noise.
		The minimum value of correlation between an acceptable stellar source and the input PSF was chosen to be 0.5 (see Sec. 3.4. in \citealt{starfinder} for more information). \tabref{tab:photinfo} shows the number of extracted sources and also the lowest mass estimated from the photometry (m$_{\rm low-obs}$) on the synthetic images. 
		Note that the noise of the sky has a different value in each synthetic image.
		As an example Sim1, Sim2 and Sim5 have same initial mass and the true number of stars in the image regions is about the same. However, a larger number of sources is extracted from Sim5 since its images have lower sky noise. Last column in \tabref{tab:photinfo} shows the SNR value for a 1$~\Msun$ star in each synthetic image.
		VLT/$\Ks$ images have higher angular resolution and better pixel-sampling than HST/V, and so more sources are detected in the VLT/$\Ks$ images compare to the HST/V images.
		The last two columns in \tabref{tab:photinfo} show the fraction of detected sources from the photometry versus both the real number of stars used to make the image (fourth column) and also the stars more massive than m$_{\rm low-obs}$ (fifth column). In all cases, less than 47\% of stars above the photometric threshold could be detected from the photometry.  
		For one of the simulations (Sim1) we found the common sources between two sets of data in HST/F555W and IRDIS/$\Ks$ filters.
		\figref{fig:cmd} shows the Color Magnitude Diagram (CMD) for these common sources.
		Among 3087 detected sources in Ks and 1322 in F555W, we could find 747 common sources. Solid lines in this figure are the {\sc{parsec}} isochrones at 1, 2, 3, 4 and 5 Myrs.
		The 2 Myr isochrone fits well with the CMD of the observed data. The conjunction of Pre-main sequence and main sequence stars is the best area to fit the isochrone, for young star clusters which does not have horizontal branches (from evolved stars) at the upper part of the CMD.
		
		We used {\sc{parsec}} isochrone at 2 Myr to estimate the stellar masses. 
		We considered a photometric error of $\sigma_{\magnitude}=0.2\magnitude$ on the apparent magnitude of the detected sources. This corresponds to the average flux error of the extracted sources and provides us with an error on the stellar masses ($\sigma_m$). 
		The uncertainty in the mass of each star was accounted for when constructing the MF.
		We estimated the slope of the MF ($\Gamma$) defined by \equref{eq:mf}, 
		
		\begin{equation}
		\logten(N) = \Gamma ~ \logten(\frac{m}{\Msun}) + {\rm constant}.
		\label{eq:mf}
		\end{equation} 
		
		where $m$ is the stellar mass and $N$ is the number of stars.
		We used an implementation of the nonlinear least-squares Marquardt-Levenberg algorithm to calculate the value of $\Gamma$ for each cluster.
		
\begin{figure}
			\includegraphics[width=\columnwidth]{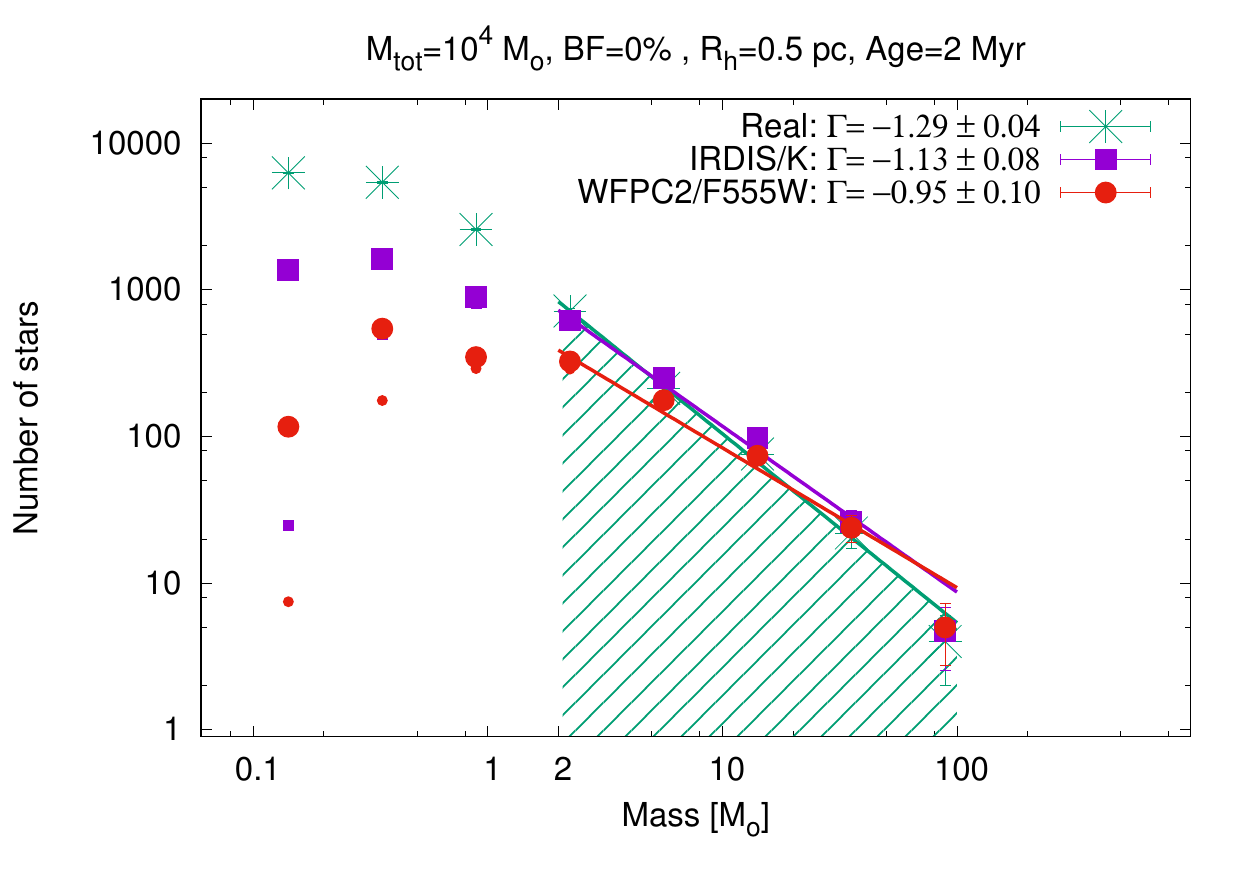}\\
			\includegraphics[width=\columnwidth]{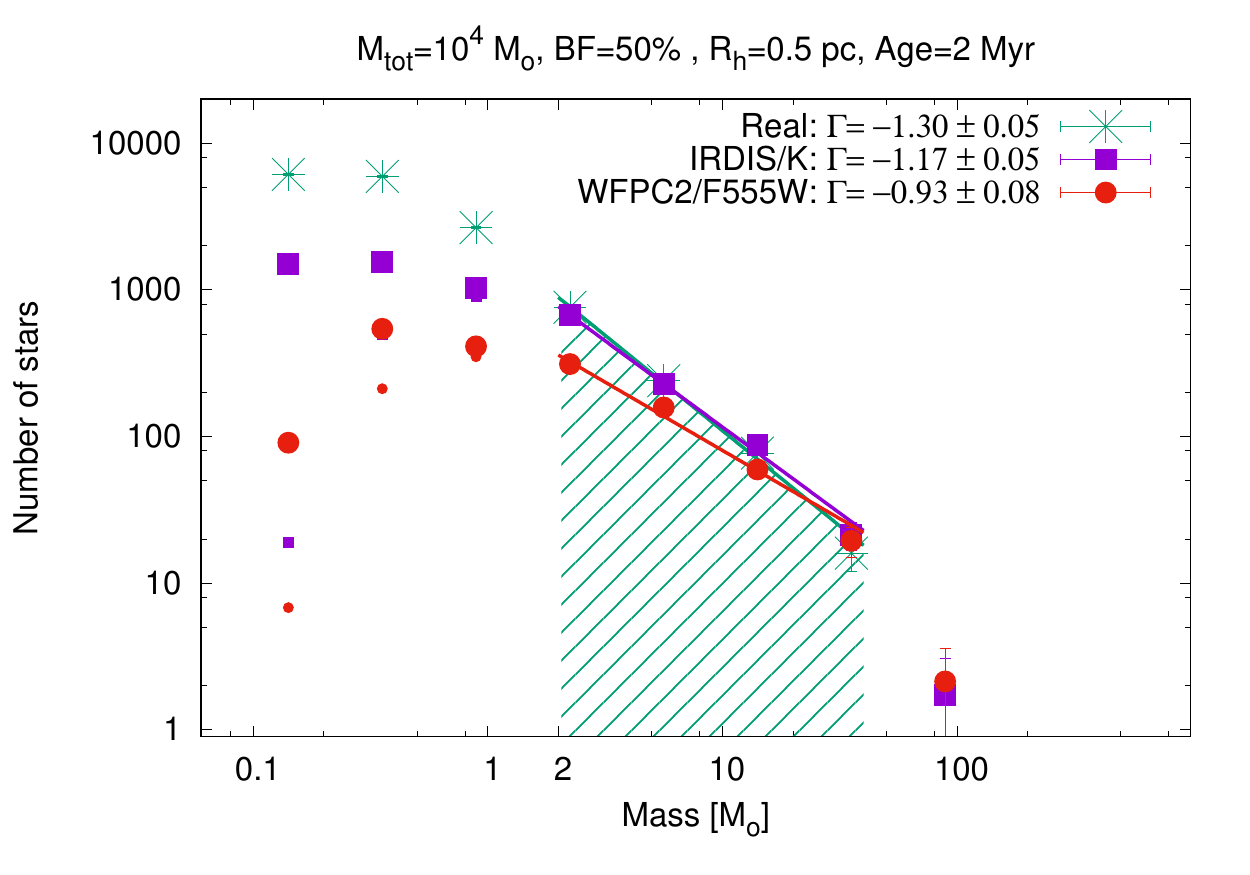}    
			\caption{MF of Sim1 (top) and Sim2 (bottom) simulations, shown in \tabref{tab:simsIC} at the age of 2\,Myr. Green is the MF derived directly from \Nbody simulations. Violet and red are the MF derived from the photometric analysis of the synthetic images created in SPHERE/IRDIS/$\Ks$ filter and HST/WFPC2/F555W filter, respectively. Both clusters are non-segregated with initial total mass of $10^4\Msun$ and a half-mass radius of $0.5$\,pc. Top has no initial binaries and bottom has 50\% initial binaries.
			Small circles/squares are the number of detected sources and large circles/squares are their completeness-corrected values.
			Green filled area shows the mass range where MF is fitted.
			}
			\label{fig:mf104}
		\end{figure}

\begin{figure}
			\includegraphics[width=\columnwidth]{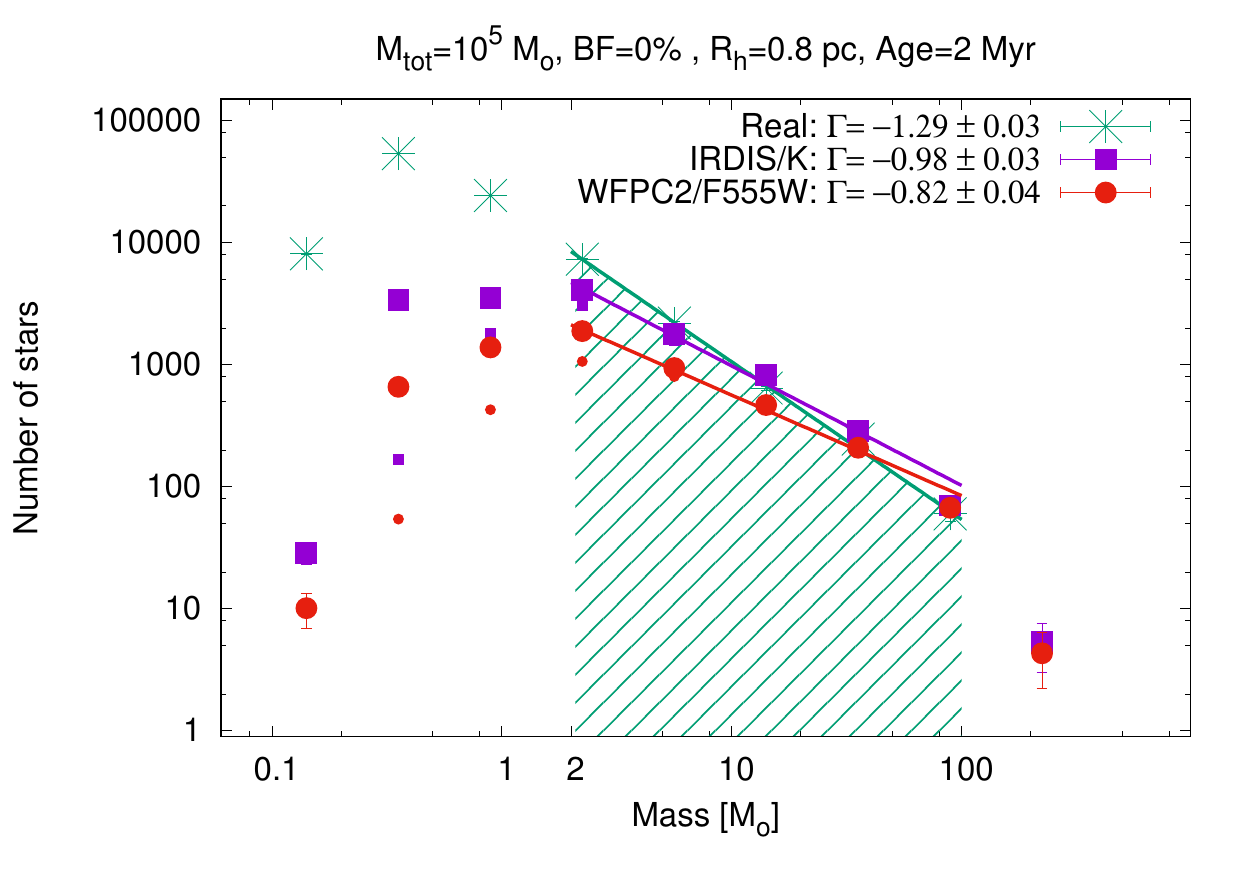}\\
			\includegraphics[width=\columnwidth]{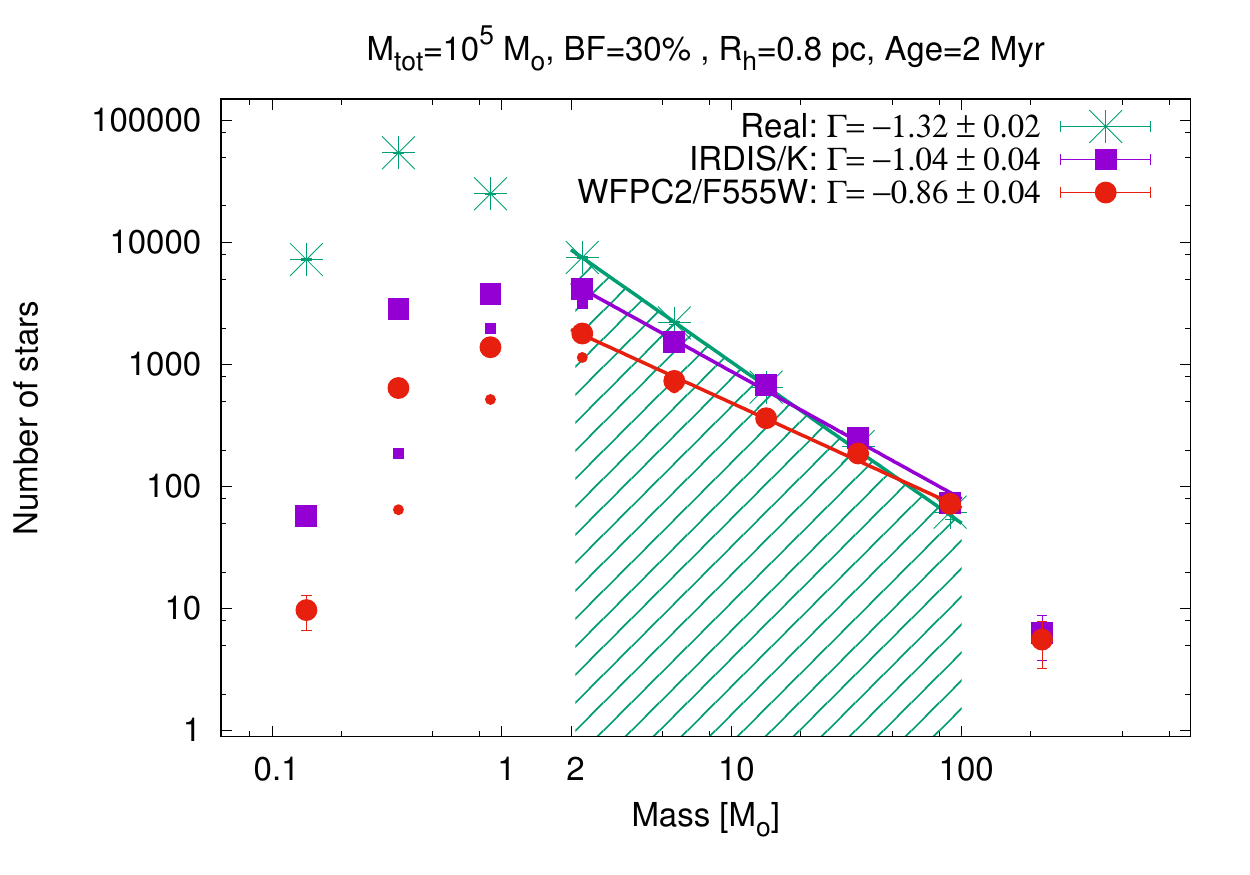}    
			\caption{Same as \figref{fig:mf104} but for Sim3 (top) and Sim4 (bottom). Both clusters are non-segregated with an initial total mass of $10^5\Msun$ and a half-mass radius of $0.8$\,pc. Top has no initial binaries and bottom has 30\% initial binaries.}
			\label{fig:mf105}
		\end{figure}

\begin{figure}
			\includegraphics[width=\columnwidth]{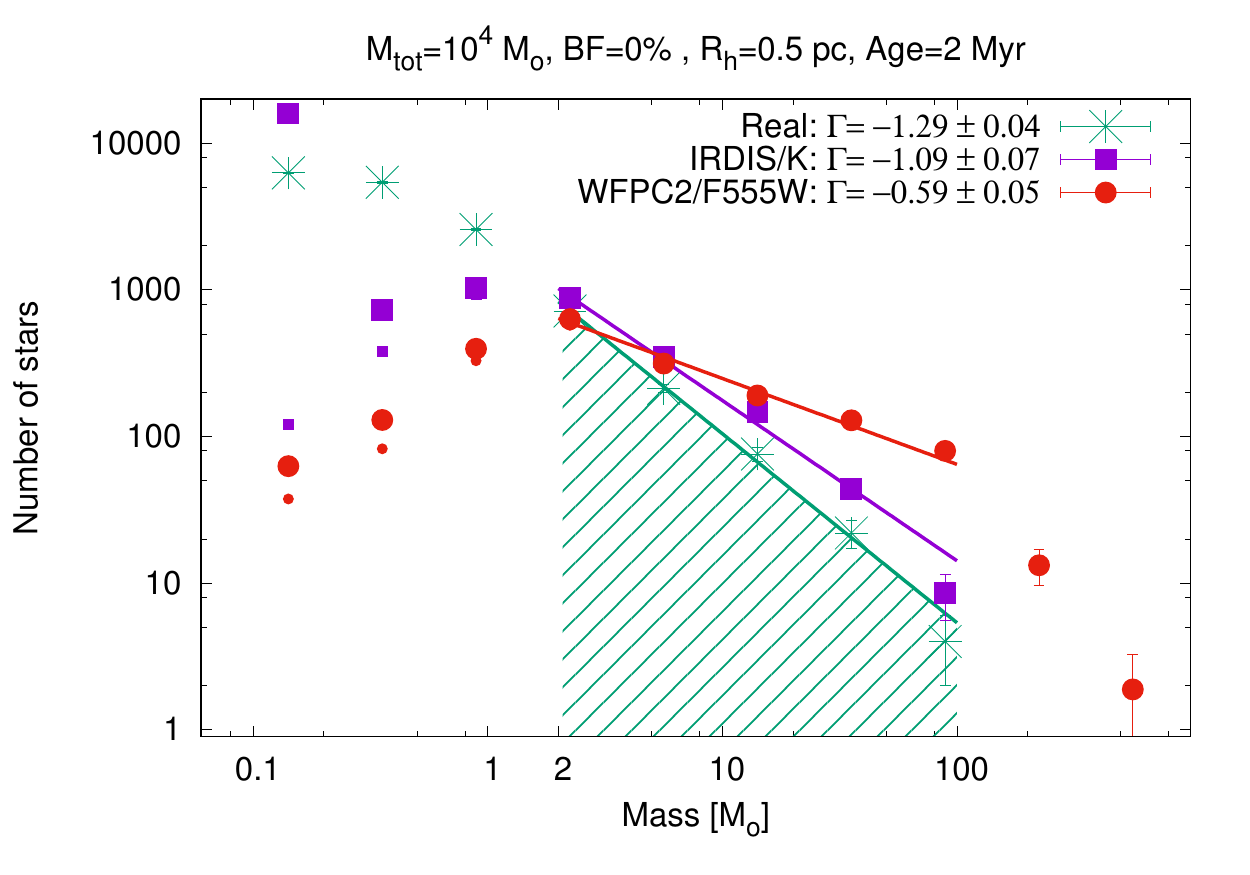}
			\caption{Same as \figref{fig:mf104} but for Sim5. The cluster is non-segregated with an initial total mass of $10^4\Msun$ and a half-mass radius of $0.5$\,pc and no initial binaries.
				This cluster contains gas with the average extinction of A$_V=3.5$.}
			\label{fig:mfext}
		\end{figure}

		We performed incompleteness tests on each synthetic image by adding artificial stars with known magnitudes, one by one, to the synthetic images. These experiments were repeated 500 times for each flux value (magnitude). 
		The completeness-corrected MF slopes are estimated for each simulation and the values are given in \tabref{tab:mfinfo}. The expected errors due to Poisson noise are considered in these fittings. 
		The low-mass limit for fitting MF is $2.0~\Msun$ and the completeness values for that mass are provided in \tabref{tab:mfinfo}. We fix the low-mass limit for all the synthetic images for the sake of the MF slopes comparison. The completeness is above 50\% at this limit in all the images.
		The high-mass limit for the MF fitting is the last point where there is a star in the underlying MF.
        Figures \ref{fig:mf104} to \ref{fig:mfext} illustrate the MFs of all the simulated clusters. Small circles/squares are the observed number of stars and large circles/squares are their completeness-corrected values.
		Green filled area shows the mass range where MF is fitted.
        One can see that the measured MFs of the simulated clusters is lower (higher) than their underlying MFs at the low-mass (high-mass) end. This is due to the fact that, at low resolution the observed flux of some of the low-mass stars falls below the detection threshold, leading to an underestimation of low-mass stars. In addition, stars which are close to each other will be counted as one for a low resolution (crowding effect), leading to an overestimation of more massive objects. 
				
		\tabref{tab:mfinfo} shows the MF slopes derived directly from \Nbody simulations ($\Gamma_{\rm real}$) and also from the photometry of the synthetic images ($\Gamma_{\Ks}$ and $\Gamma_V$). 
		The MF slopes of the clusters estimated from the synthetic images with low resolution (HST/F555W) are flatter than those with a higher resolution (SPHERE/IRDIS/$\Ks$) as well as the underlying MFs derived from the \Nbody simulations.
As explained earlier, this is due to the fact that we underestimate (overestimate) the number of low-mass (high-mass) stars as a result of low resolution. Therefore, as observational conditions become poorer, the observed MF becomes flatter, mimicking mass-segregation. 
		
		Sim5 is embedded in a homogeneous cloud and has a variable extinction (see \figref{fig:ext}) throughout the cloud as each star has a different distance from the observer. In the photometric analysis of the real observational data from star clusters, the extinction is often considered to be a constant value which is applied to all detected sources in the FOV. If we consider a mean value of A$_{V}$ (3.5\,mag) for Sim5 in our photometric analysis, the stellar mass estimation would be affected significantly. As it can be seen from \tabref{tab:mfinfo} and \figref{fig:mfext}, among all the simulated clusters, Sim5 has the shallowest observed MF. In particular, Sim5 has an underlying MF slope of $\Gamma_{\rm real}= -1.29 \pm 0.04$ and $\Gamma_V = -0.59 \pm 0.05$ from HST/WFPC2/F555W images. The is because we take the mean value of A$_V$ for all the stars. As a result, some objects will have an overestimated (underestimated) A$_V$ making them over-luminous (under-luminous) and thus more (less) massive, explaining the very different shape of the MF for sim5. Given the fact that Sim5 has the same observational conditions as Sim1 with the addition of extinction, and that the MF slope of Sim5 is $\sim0.4$\,dex shallower than that of Sim1 in the visible, indicates that extinction has a major effect on the measured MF slopes of YSCs. 
		
		According to \tabref{tab:mfinfo}, binaries do not affect the high-mass slope of the MF significantly. 
		Note that the initial pairing of binary systems can affect the MF in addition to the observational biases. 
		
		\begin{table*}
			\centering
			\caption{The observed number of stars (N$_{\rm obs}$) and the lowest mass observed (m$_{\rm low-obs}$) in the photometric analysis.
			$\frac{\rm N_{\rm obs}}{\rm N_{\rm total}}$ shows the fraction of observed sources versus real number of stars used to create the synthetic image.
			$\frac{\rm N_{\rm obs}}{\rm N_{(\rm m>\rm m_{\rm low-obs})}}$ shows the fraction of observed sources versus the real number of stars more massive than  m$_{\rm low-obs}$.
			SNR is the signal to noise ratio of a 1 $\Msun$ star in the FOV of each image. For Sim5, the average value of extinction (A$_V=3.53$) is considered in front of the 1 $\Msun$ star.
}
			\label{tab:photinfo}
			\begin{tabular}{lccccc}
				\hline
				ID/Telescope/Filter & N$_{\rm obs}$ & m$_{\rm{low-obs}}$ &$\frac{\rm N_{\rm obs}}{\rm N_{\rm total}}$& $\frac{\rm N_{\rm obs}}{\rm N_{(\rm m>\rm m_{\rm low-obs})}}$&SNR\\
				 &  & [$\Msun$] && &for 1 $\Msun$ \\
				\hline
				Sim1/VLT/$\Ks$& 3087 & $0.33^{+0.20}_{-0.10}$&0.202&0.462&12.46\\
				Sim1/HST/V& 1322 & $0.33^{+0.11}_{-0.10}$&0.086&0.198&74.47\\
				Sim2/VLT/$\Ks$& 3268 & $0.33^{+0.20}_{-0.10}$&0.207&0.460&12.46\\
				Sim2/HST/V   & 1433 & $0.33^{+0.11}_{-0.09}$&0.091&0.202&74.47\\
				Sim3/VLT/$\Ks$& 9842 & $0.68^{+0.15}_{-0.03}$&0.102& 0.357&8.07\\
				Sim3/HST/V   & 3558 & $0.53^{+0.12}_{-0.00}$&0.037& 0.093&26.18\\
				Sim4/VLT/$\Ks$& 10007 & $0.68^{+0.15}_{-0.03}$&0.102& 0.351&8.07\\
				Sim4/HST/V   & 3586 & $0.55^{+0.10}_{-0.02}$&0.037& 0.095&26.18\\       
				Sim5/VLT/$\Ks$& 3746 & $0.16^{+0.55}_{-0.01}$&0.245& 0.320&25.95\\
				Sim5/HST/V   & 2119 & $0.12^{+1.05}_{-0.01}$&0.138& 0.150&905.48\\       
				\hline
			\end{tabular}
		\end{table*}

		\begin{table}
			\centering
			\caption{MF slope ($\Gamma$) given in \equref{eq:mf} for the simulated clusters. $\Gamma_{\rm real}$: MF slopes derived directly from \Nbody simulations. $\Gamma_{\Ks}$, $\Gamma_{V}$: MF slopes from the photometric analysis on the synthetic images of SPHERE/IRDIS/$\Ks$ and HST/WFPC2/F555W, respectively. The low-mass limit for MF fitting is 2$\Msun$ and C is the completeness value at this limit.
            The MF slopes are corrected for completeness and fitted with expected errors due to Poisson noise.}
			\label{tab:mfinfo}
			\begin{tabular}{lccccc} 
				\hline
				ID & $\Gamma_{\rm real}$ & $\Gamma_{\Ks}$ &C[\%] &$\Gamma_{V}$& C[\%]\\
				\hline
				Sim1 & $-1.29 \pm 0.04$ & $-1.13 \pm 0.08$ &96& $-0.95 \pm 0.10$&88\\
				Sim2 & $-1.30 \pm 0.05$ & $-1.17 \pm 0.05$ &95& $-0.93 \pm 0.08$&94\\
				Sim3 & $-1.29 \pm 0.03$ & $-0.98 \pm 0.03$ &74& $-0.82 \pm 0.04$&56\\
				Sim4 & $-1.32 \pm 0.02$ & $-1.04 \pm 0.04$ &75& $-0.86 \pm 0.04$&63\\
				Sim5 & $-1.29 \pm 0.04$ & $-1.09 \pm 0.07$ &96& $-0.59 \pm 0.05$&87\\
				\hline
				\end{tabular}
		\end{table}

		\subsection{Spectroscopy}\label{sec:spectroscopy}
		
\begin{figure}
			\includegraphics[trim=20 0 45 0,clip,width=\columnwidth]{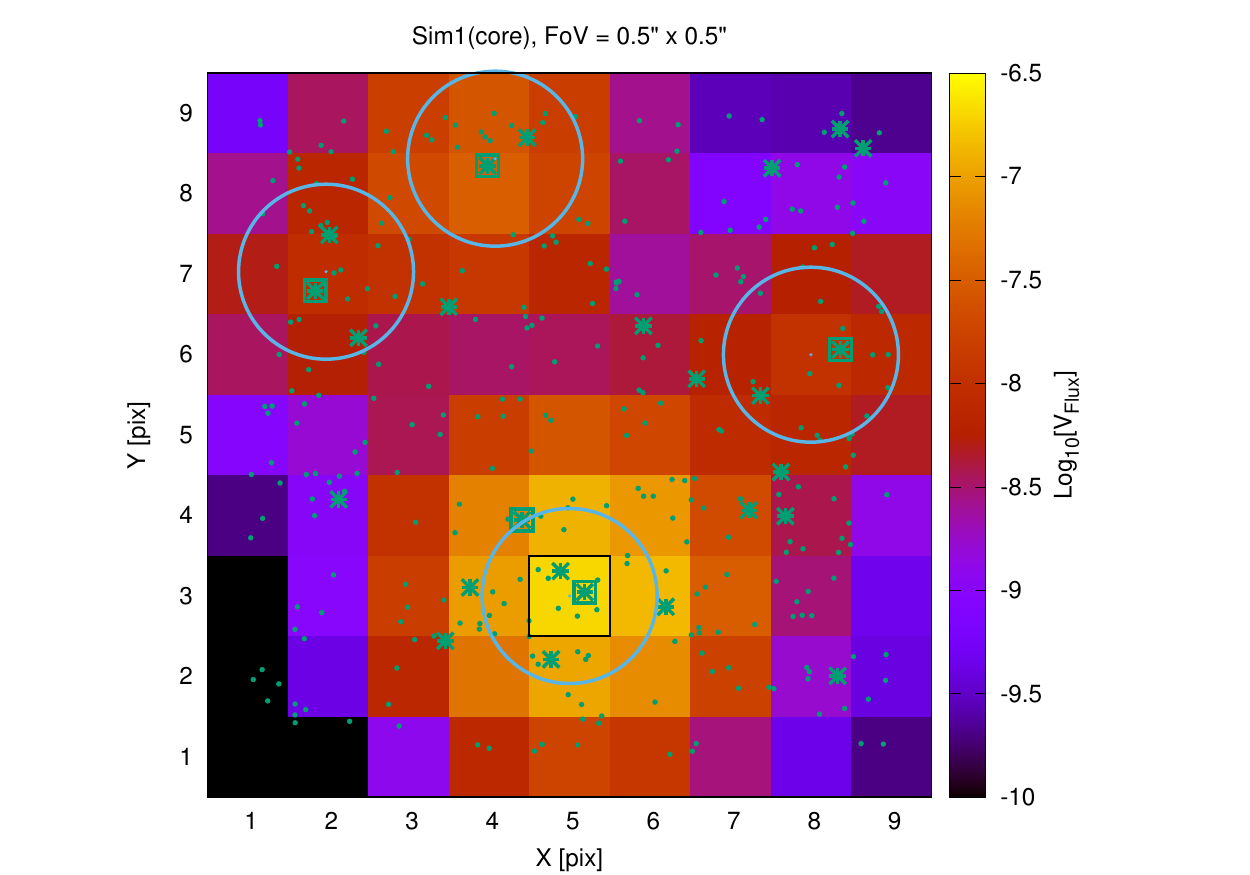}	\includegraphics[width=\columnwidth]{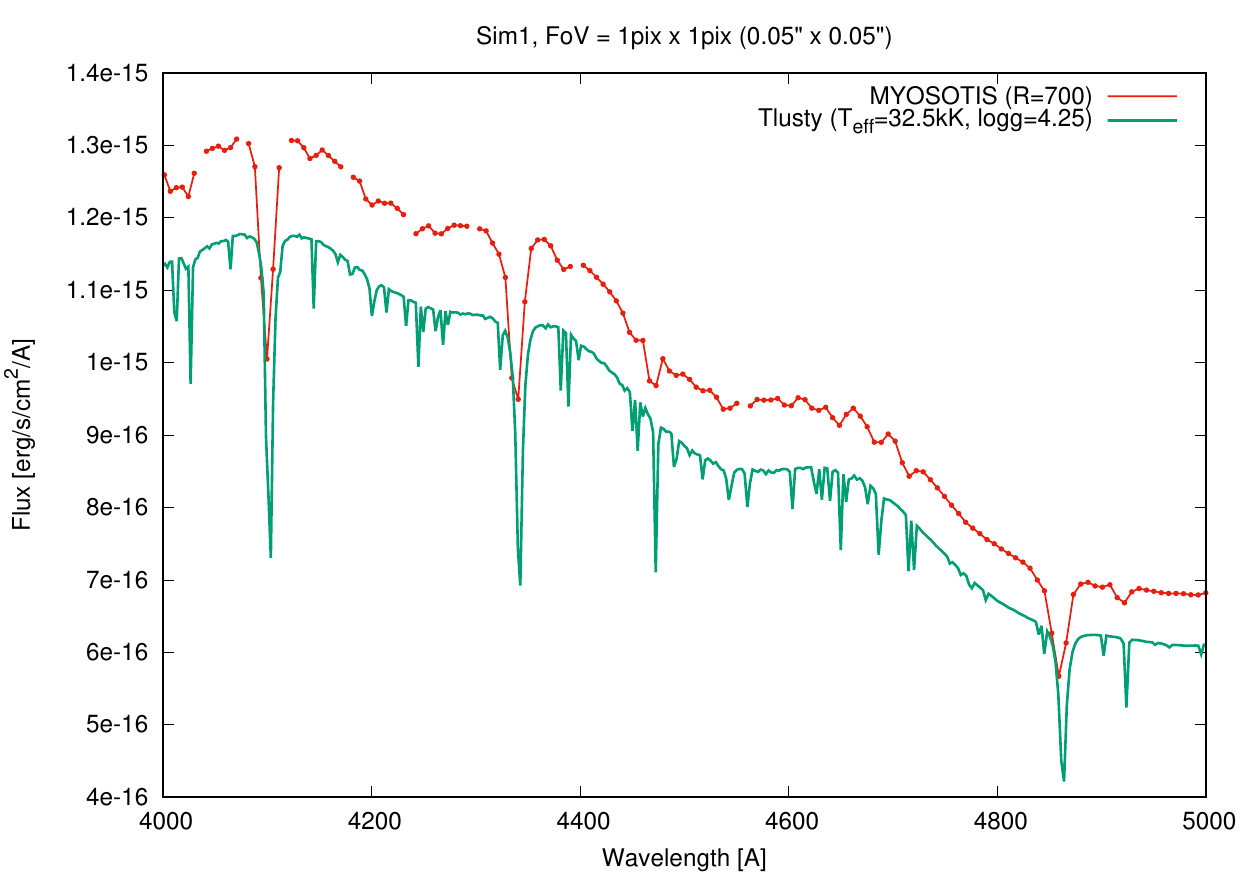}
			\caption{Top: Center of Sim1 star cluster (FoV=$0.5"\times0.5"$). MYOSOTIS created a cube of this region so that the spectra along each pixel is available. Stars with masses bellow $1  \Msun$ are shown with a small dots. $1.0-3.5 \Msun$ stars are shown with the star-signs. The OB stars ($T_{\rm eff} >$ 15 kK) are shown with the green squares. Large blue circles shows stars detected from the photometric analysis.
				Bottom: shows the example of spectra along the black pixel where the most massive star in the FoV is located. Red is a low resolution spectra created by \myosotis and Green is a {\sc{tlusty}} SED for the represented star at the distance of 50\,kpc in HST/F450W filter.
			}
			\label{fig:smallfov}
		\end{figure}

		\figref{fig:smallfov} shows the synthetic spectroscopic data produced by \myosotis from the center of Sim1 cluster. \myosotis created a cube of this region (FOV of  $0.5" \times 0.5"$) so that the spectra along each pixel is available. The image has the angular resolution of HST/WFPC2 ($\sim 0.110"$) in the HST/F450W filter and the spectroscopic resolution (${\lambda}/{\Delta \lambda}$) of 700 covering wavelengths $(3900-5100)~\textup{\AA}$. Creating the spectroscopic cube is computationally expensive so as an example we just created a small FOV with low spectral resolution data.
	The Doppler shift on the star's spectra from their velocity were not applied (although this is a feature in \myosotis), so that we can demonstrate the affect of pure blending on the observational data.
		
		In this small region there are 327 stars (green dots in \figref{fig:smallfov}). 25 stars have masses above 1 $\Msun$ (green stars in \figref{fig:smallfov}) and 5 of them are O- and B-type stars (green squares in \figref{fig:smallfov}) with $T_{\rm eff} >$ higher than 15000K.
		The large blue circles in \figref{fig:smallfov} shows the detected sources from the photometric analysis. 
		\tabref{tab:smallfov} shows the magnitude and position (first and second columns) of B- and O-type stars in this FOV, compared to the magnitude and position derived from photometric analysis (third and forth columns). One can see how poorly an observer can detect stellar sources. Among 327 stars just 4 of them are detected. None of the medium- and low-mass stars are detected in this region, and we have trouble detecting even the close-by massive stars.
        
We plotted an example of the spectra on the pixel highlighted with the black square. This is a pixel where the most massive star, in this FOV, is located (first star in \tabref{tab:smallfov}). 
 The comparison between low-resolution observed flux created by \myosotis and the SED of the brightst star ($T_{\rm eff}=32500\rm K$ and log$\,$g$=4.25$) from {\sc{tlusty}} is shown in \figref{fig:smallfov}. This SED is chosen by \myosotis according to the metallicity, mass and age of this star (see \secref{sec:EAmodels} on how \myosotis chooses the proper SED for stars). The green line in \figref{fig:smallfov} shows the SED of the brightest star observed at the distance of 50\,kpc and multiplied by the chosen filter transparency.

The observed flux is higher than that of the brightest star's intrinsic flux, and the spectrum also has a different shape. This is because the observed flux is blended with the nearby detected sources, as well as an undetected early B-type star) and numerous undetected medium- and low-mass stars. This demonstrates how \myosotis can be used to examine the spectroscopic properties of stellar systems in clusters.

		
		\begin{table}
			\centering
			\caption{Magnitude and position of B- and O-type stars in \figref{fig:smallfov} are shown in the first and second columns.
				The photometric magnitude and positions (large blue circles in \figref{fig:smallfov}) are given in the third and forth columns.
			}
			\label{tab:smallfov}
			\begin{tabular}{cc|cc} 
				\hline
				mag$_{F555W}$ & X[pix],Y[pix]& mag$_{phot}$ & X[pix],Y[pix]  \\
				\hline
				14.8956    &    5.1893, 3.0490 	&14.894 	&	5.000, 3.000   \\
				16.9248    &    3.9822, 8.3474 	&16.877 	&	4.075, 8.436 \\
				18.3018     &   4.4093, 3.9458 	&	-			&	-	\\
				18.3018    &   8.3662, 6.0624 	&18.561  &	8.000, 6.000\\
				18.5249    &   1.8390, 6.7939  	&18.435 	&	1.974, 7.032	\\
				\hline
			\end{tabular}
		\end{table}
		
		\section{Summary and Conclusions}\label{sec:summary}
		In this paper we introduced \myosotis (Make Your Own Synthetic ObservaTIonS), an IDL {\footnote{The IDL version of the source code of \myosotis is publicly accessible in \url{https://github.com/zkhorrami/MYOSOTIS}, and a Python version is currently under development.}} code written tool for simulating synthetic observation for space or ground based telescopes which can synthesis both imaging and spectroscopic data. \myosotis can generate the synthetic images from user input (given the position, age, and the mass of stars as well as the extinction values for the FOV) or from the output of \Nbody simulations (e.g. \nbodysix) as well as SPH simulations (e.g. \gandalf). It uses the {\sc{parsec}} evolutionary models and different atmosphere models to estimate the flux of stellar sources in different filters within a given FOV (see \secref{sec:EAmodels}). The observing conditions, instrumental resolution and noise can be specified by the user (\secref{sec:myosotis}). \myosotis is a highly customizable tool, with the user being able to define their own input models, such as filters, models for stellar evolution, stellar atmospheres, etc. The \myosotis library of SEDs and evolutionary models can also be replaced easily to the updated models by the user. For example the high resolution spectroscopic data from GAIA can be used as an input for different spectral types stars.
		
		As an example of the application of \myosotis in this paper, we created synthetic HST/WFPC2/F555W and VLT/SPHERE/IRDIS/$\Ks$ images of five YSCs at the age of 2\,Myr with Solar metallicity, in the visible and nearIR. Each cluster had a different initial total mass, half-mass radius, binary fraction, and extinction. 
		Photometry on each image was done using {\sc{starfinder}} package. We used {\sc{parsec}} evolutionary models for re-estimating cluster's age by isochrone fitting on the CMD. 
		The stellar masses are re-estimated using 2Myr {\sc{parsec}} isochrone and the average value of the extinction in the FOV.
        The underlying MF and the observed MF of each cluster (subject to different observing conditions) was derived by considering the error on stellar masses. We also performed the artificial star test on each synthetic image to estimate the completeness-corrected MF slopes (\tabref{tab:mfinfo}).
		In all the cases, the slope of the MF ($\Gamma$) becomes flatter as the resolution decreases. All MF slopes of the HST synthetic images are flatter than those of SPHERE, which in turn are both flatter than the underlying MFs of YSCs obtained directly from \Nbody simulations.  Standard completeness tests do not seem to help here. This is likely because they assume that stars are randomly positioned in the stellar field, while in clusters such as those we examine here, there is significant sub-clustering (even after 2 Myr of dynamical N-body evolution). This sub-clustering is not taken into account in the completeness tests, when randomly positioning the fake sources, and we so we tend to overestimate the completeness.   
		
		Moreover, according to our analysis, the difference between the measured MF of clusters with an initial binary population and those without binaries is $\leq0.1$\,dex, i.e. the effect of binaries on the high-mass slope of the MF is marginal. This study suggests that the observed discrepancy in the reported values of the MF slopes of YSCs (such as R136 and NGC 3603) could primarily be due to how incompleteness is treated, rather than the unresolved binary population. However significantly more work needs to be done to see whether this is indeed the case. We aim to explore this in an upcoming paper.
		
		Our analysis also confirms that extinction can have a major effect on the measured value of the MF slopes, especially at visible band wavelengths. In particular, we demonstrated that in the presence of extinction the measured MF slope of a YSC in HST/WFPC2/F555W images is $\sim0.4$\,dex shallower compared to when there is no extinction (compare Sim5 with Sim1 in \tabref{tab:mfinfo}). 
		The affect of resolution on both photometry and spectroscopy data can be seen in details in the example of spectrsocopic-image cube from the center of Sim1 (\figref{fig:smallfov}-top). Only 16\% (0.01\%) of stars with masses above 1 $\Msun$ (0.1 $\Msun$) is detected by photometry. The spectra of a typical O-type star is blended with the nearby undetected sources (\figref{fig:smallfov}-bottom). 
	This result suggests that the observed mass segregation reported for some young star clusters (see MF slopes reported in \cite{Eisenhauer98,Sung04,Stolte06,Harayama08,pang,khorrami16} for NGC3603 and \cite{Malumuth94,hunter96,massey98,Brandl96,Selman99,sirianni2000,andersen2009,khorrami17} for R136), could also be explained by observational confusion for the lack of angular resolution and unknown extinction values across the observed FOV. However more work would need to be done to support this conclusion.
		
		In principle, the shape of MF for a given simulated cluster can be affected by binaries, multiple populations, 
        patchy extinction,
        telescope limitations,
        and incompleteness estimates. 
        \myosotis can be used to investigate the effect of the above-mentioned parameters on the MF estimated from different observational instruments. It can also be used to explore the effects of age-determination. For example, a wrong estimation of the age of the star cluster, which can be caused by multiple stellar populations or extinction in very broad observing filters (e.g. Gaia g filter), will also also have an affect on the derived MF. \myosotis is also useful for projects such as studying multiple populations in star clusters, and their affect on observationally derived parameters. This could also be used to look at cluster merger events, where each cluster has different ages and/or metallicities. The physical properties of the binary (or multiple) systems measured from synthetic observations, can be compared with their original values from the simulations.

        It should be stressed that for simulating clusters with extinction, \myosotis is designed to look at the visible and near infrared wavelengths where dust re-emission is not significant.  When this approximation fails, one needs to resort to a more detailed radiative transfer approach (e.g. \citealt{Koepferl2017}).

		\section*{Acknowledgements}
		We would like to thank the reviewer for their constructive and insightful comments. The StarFormMapper project has received funding from the European Union's Horizon 2020 research and innovation programme under grant agreement No 687528.
		This research has made use of the SVO Filter Profile Service ({\url{http://svo2.cab.inta-csic.es/theory/fps/}}) supported from the Spanish MINECO through grant AyA2014-55216. PCC acknowledges support from the Science and Technology Facilities Council (under grant ST/N00706/1).
		Finally, we would also like to thank Lee G. Mundy for his suggestions which led to the improvement of this work.
		

		\bsp	
		\label{lastpage}
	\end{document}